\documentclass[twocolumn]{aastex7}
\usepackage{amsmath,amssymb,amsfonts}
\usepackage{natbib}
\usepackage{threeparttable}
\usepackage{multirow}
\usepackage{subfigure}
\usepackage{graphicx}
\usepackage[OT1,T1]{fontenc}

\DeclareRobustCommand{\ion}[2]{\textup{#1\,\textsc{\lowercase{#2}}}} 

\def\CaIR{\ion{Ca}{ii}~8542\,\AA} 
\def\FeIline{\ion{Fe}{i}~6173\,\AA} 
\def\Ca3{\ion{Ca}{ii}}

\begin{document}

\title[Stokes Profile of Chromospheric Reconnection]{Synthetic \CaIR\ Stokes Profile Associated with Chromospheric Magnetic Reconnection in a Simulated Active Region}

\author{Xinyu Zhou}
\affiliation{Department of Earth and Planetary Science, The University of Tokyo, 7-3-1 Hongo, Bunkyo-ku, Tokyo, 113-0033, Japan}
\email{zhouxy@eps.s.u-tokyo.ac.jp}

\author{Takaaki Yokoyama}
\affiliation{Astronomical Observatory, Kyoto University, Sakyo-ku, Kyoto 606-8502, Japan}
\email{yokoyama.takaaki.2a@kyoto-u.ac.jp}

\author{Haruhisa Iijima}
\affiliation{Centre for Integrated Data Science, Institute for Space-Earth Environmental Research, Nagoya University, Furocho, Chikusa-ku, Nagoya, Aichi 464-8601, Japan}
\email{h.iijima@isee.nagoya-u.ac.jp}

\author{Takuma Matsumoto}
\affiliation{Centre for Integrated Data Science, Institute for Space-Earth Environmental Research, Nagoya University, Furocho, Chikusa-ku, Nagoya, Aichi 464-8601, Japan}
\affiliation{National Astronomical Observatory of Japan, 2-21-1  Osawa, Mitaka, Tokyo, 181-8588, Japan}
\email{takuma.matsumoto@gmail.com}

\author{Shin Toriumi}
\affiliation{Institute of Space and Astronautical Science, Japan Aerospace Exploration Agency, 3-1-1 Yoshinodai, Chuo-ku, Sagamihara, Kanagawa 252-5210, Japan}
\email{toriumi.shin@jaxa.jp}

\author{Yukio Katsukawa}
\affiliation{National Astronomical Observatory of Japan, 2-21-1  Osawa, Mitaka, Tokyo, 181-8588, Japan}
\email{yukio.katsukawa@nao.ac.jp}

\author{Masahito Kubo}
\affiliation{National Astronomical Observatory of Japan, 2-21-1  Osawa, Mitaka, Tokyo, 181-8588, Japan}
\email{masahito.kubo@nao.ac.jp}

\begin{abstract}
Magnetic reconnection is an important driving mechanism of many chromospheric phenomena, e.g., UV bursts and chromospheric jets. Information about magnetic field is indispensable for analyzing chromospheric magnetic reconnection, which is mainly encoded in polarization signals. The purpose of this work is to predict possible Stokes features related to chromospheric reconnection events, from realistic two-dimensional magnetohydrodynamic simulation and Stokes profile synthesis. An emerging magnetic flux sheet is imposed at the bottom boundary of a well-relaxed unipolar atmosphere that spans from the upper convection zone to the corona. The reconnection region is heated to $\sim7$~kK and the outflow velocity reaches up to $\sim35$~km~s$^{-1}$. Through Stokes profile synthesis, several Stokes features related to reconnections and plasmoids are reproduced. We found sign reversal features on circular polarization and amplitude reduction features on linear polarization at reconnection sites. Also, we report strong linear and circular polarization signals corresponding to huge ($\sim300$~km) and tiny ($\sim40$~km) plasmoids, respectively. We conclude that both linear and circular polarization signals may reveal the distinctive physical mechanisms in reconnections, and enhance the understanding of magnetic reconnection in observations.
\end{abstract}


\section{Introduction} \label{sec:intro}
The solar chromosphere falls in between the cool but high-density photosphere and the hot but low-density corona. While the dynamics of the high plasma $\beta$ photosphere are dominated by gas motions, the low plasma $\beta$ corona is mainly governed by magnetic force. As the intermediate layer, the solar chromosphere is under the joint control of these two effects, which enable complex evolution processes. Especially, with much more energy input, emerging flux region, which is the early state of the active region when flux emerges from below, is linked to many dramatic phenomena in the solar chromosphere, e.g., Ellerman Bombs, UV Bursts, and chromospheric jets \citep{Ellerman1917,Peter2014,Chae99,Shimizu15,Toriumi15,Toriumi17}. Many of these chromospheric phenomena could be explained as the consequences of magnetic reconnection \citep{Innes1997,Shibata2007,Vissers2015}.

Magnetic reconnection is a ubiquitous physical mechanism in magnetized plasma. It enables the rapid transfer of magnetic energy into kinetic energy and internal energy within a short timescale. Thus, it is considered as one of the physical explanations for many dramatic phenomena in the solar atmosphere \citep{Kopp1976,Pontin2022}. Previous studies have reproduced various kinds of magnetic reconnection in the solar chromosphere \citep{ni2015,ni2018}. Besides, flux emergence has been well studied as the trigger mechanism of chromospheric magnetic reconnection \citep{Isobe2007,Takasao2013,NS2016,Hansteen2017}, which could be regarded as the evolution dynamics of an active region. With the help of spectrum synthesis, the realistic simulated reconnection events have been compared to observation results \citep{NS2017,Hansteen2019,ni2021}. These studies have extensively discussed temperature and outflow velocity in relation to spectrum intensity. However, as the essential part of magnetic reconnection, local magnetic field has not been sufficiently investigated with observational signals before.

In observations, information about magnetic field is primarily encoded in the polarization signal. Photospheric lines such as \FeIline\ and 6302\,\AA\ have been widely used for magnetic field diagnosis in previous studies \cite[e.g.,][]{Stenflo73,Schrijver03,Tsuneta08,Scherrer12}. However, difficulties arise when adapting similar approaches to chromospheric lines. Under typical local field strengths, chromospheric lines exhibit lower contrast between the spectral line widths and the Zeeman splitting widths. This difference, coupled with the lower Land\'e factors, results in lower sensitivity to polarization signals in chromospheric lines. Also, local thermodynamic equilibrium (LTE) assumption, which could decrease considerable computational cost, cannot be adapted to the synthesis process in the non-LTE chromosphere. Synthesizing most of the chromospheric lines appropriately requires partial frequency redistribution (PRD) for the photon scattering mechanism. Nevertheless, for \Ca3\ triplet infrared (IR) lines, complete frequency redistribution (CRD) leads to similar result to PRD, which could highly reduce the computational cost \citep{uiten1989}. Among the three \Ca3\ IR lines, \CaIR\ shows the highest polarimetric sensitivity. Thus, it is regarded as one of the best candidates for chromospheric magnetic field diagnosis. 

Continuous efforts have been devoted to exploring the potential of \CaIR\ line for chromospheric diagnosis, from early studies such as \citet{SN06} and \citet{Pietarila07}, to more recent works \citep[e.g.,][]{QN2016,Leenaarts18}. The core part of the spectrum is dominantly under the control of the physical property of the chromosphere. Also, the core parts of polarization signal, i.e., the Stokes~$Q$, $U$, $V$ signal of full Stokes profile, mainly depend on the magnetic field. Specifically, for magnetic field strength $|B|<1000$~G, linear polarization signals, i.e., Stokes~$Q$, $U$, have positive correlation with transverse magnetic field; while circular polarization signal, Stokes~$V$, is proportional to longitudinal magnetic field \citep{ctno2018}. Thus one could expect strong correlation between magnetic fields and polarization signals, which facilitates the diagnostic of magnetic field properties in observations. This potential has further inspired significant interest in theoretical modeling of full Stokes profiles \cite[e.g.,][]{QN2019,Matsumoto23,Kawabata24}.

Previous observational studies \cite[e.g.,][]{Yang14,Joshi20,RvdV23,Singh24} have traditionally speculated chromospheric reconnection events mainly based on spectral brightening, Doppler shifts, and the concentration of photospheric bipolar fields. However, similar features can also appear due to shocks and other unrelated phenomena. This ambiguity indicates a clear need for deeper information to confidently identify reconnection events. Recent advances on Stokes inversion techniques allow observational studies \cite[e.g.,][]{Vissers19,DB2021,daSS22} to retrieve more detailed atmospheric and magnetic field information. Yet, the limited number of nodes in the inversion process may result in the oversimplification of the reconnection details. While inversions offer a more rigorous analysis, they may still fall short in capturing the full complexity of reconnection dynamics. To supplement the current observational and inversion analyses, this study employs two-dimensional magnetohydrodynamic (2D MHD) numerical modeling to predict the characteristics of \CaIR\ Stokes features associated with chromospheric magnetic reconnection, and to connect these features with their underlying physical mechanisms. This paper is organized as follows: Section \ref{sec:mtd} describes the details of the simulation setup and the synthesis process of the Stokes profile. In section \ref{sec:res}, we analyzed the magnetic reconnection process of two chromspheric events and discussed the Stokes features which appears during the process. Section \ref{sec:dis} consists of two parts. In the first part, we compared our result with the observational capabilities and previous studies. In the latter part, we investigated the Stokes features of shock waves and summarized the main differences in the Stokes feature between magnetic reconnections and shock waves. Finally, Section \ref{sec:ccls} contains a summary.

\begin{figure*}[!t]
  \centering
  \includegraphics[width=\textwidth]{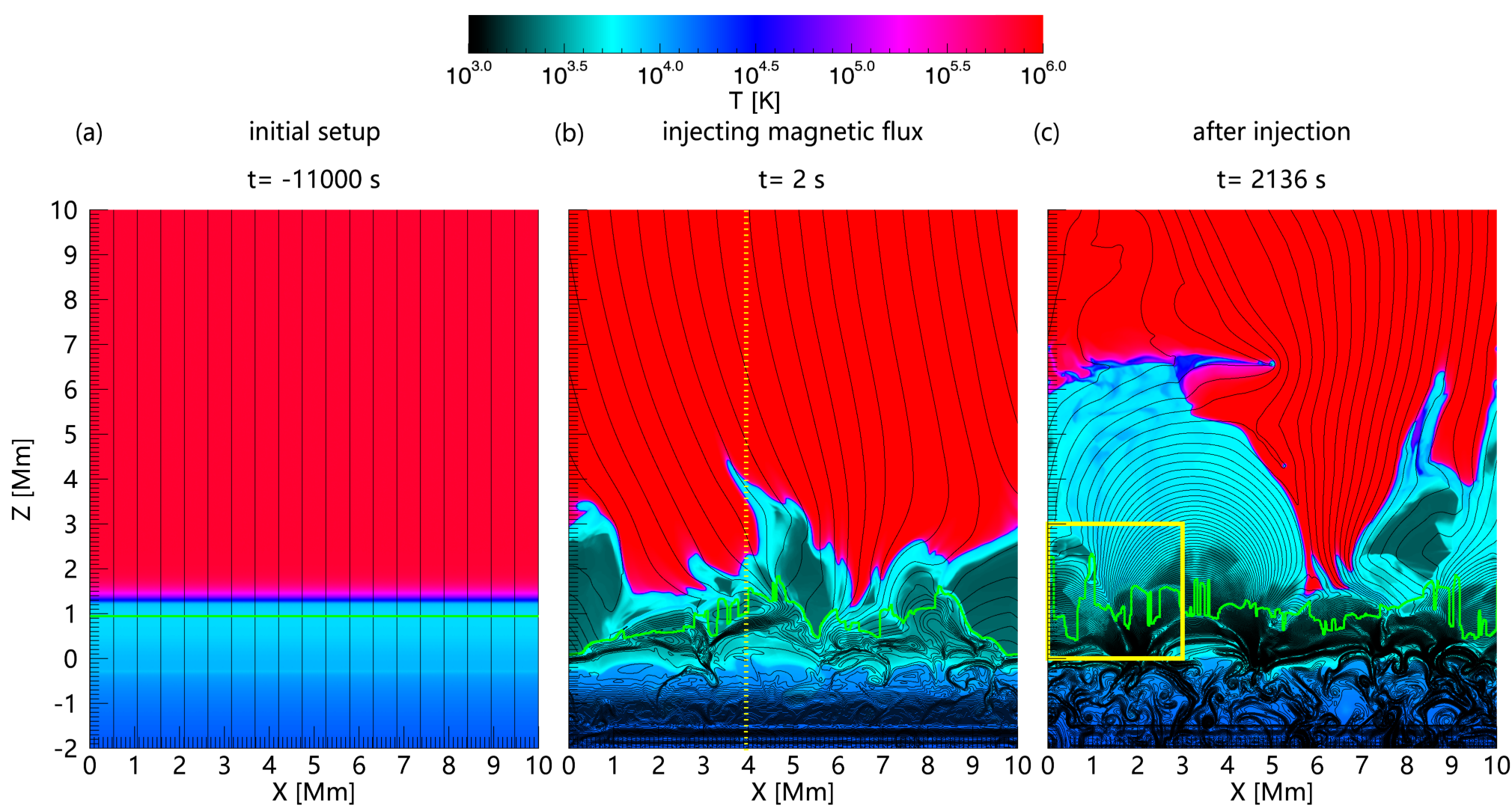}
  \caption{Temperature distribution at different stage: (a) initial setup; (b) magnetic flux imposed in the convection zone; (c) magnetic flux emerged to the chromosphere. The black contour lines indicate the magnetic field lines. The lime contour lines indicate the mean optical depth $\tau=1$ surface in the wavelength range $\lambda\sim\lambda_0\pm0.25$~\AA.}
  \label{fig:sim3snap}
\end{figure*}

\begin{figure}
  \centering
  \includegraphics[width=9.0cm]{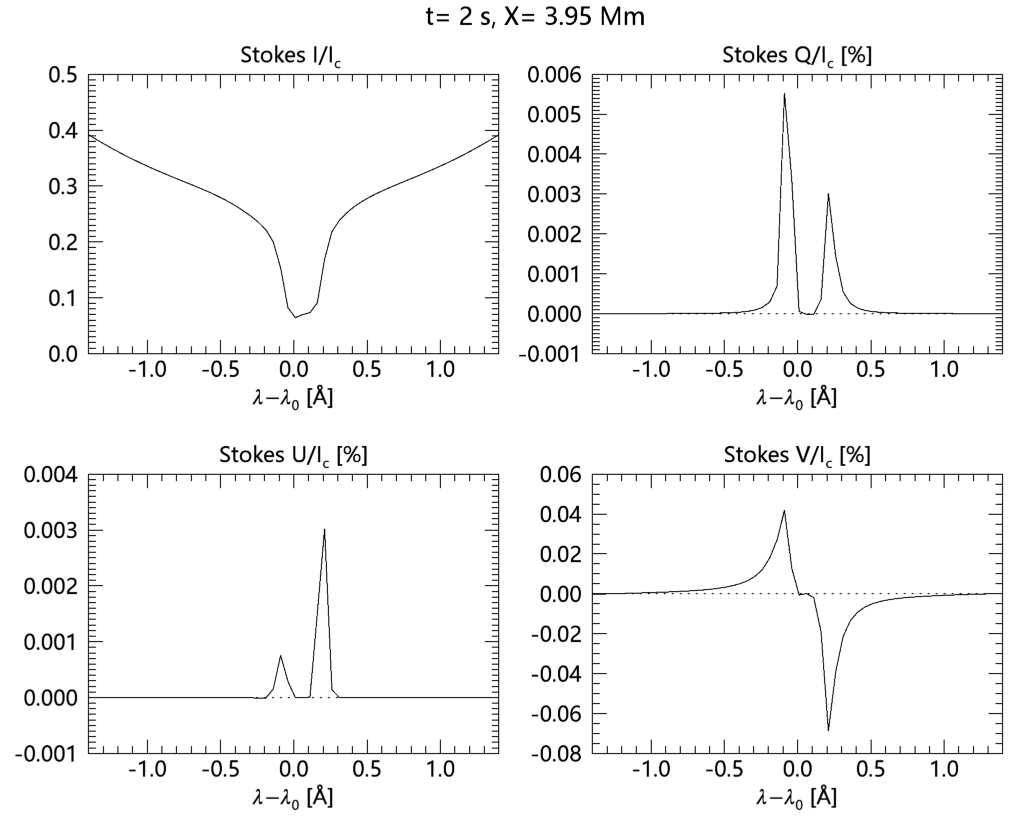}
  \caption{An example of synthetic Stokes profile at $x=3.95$~Mm from the snapshot at $t=2$~s (yellow dotted line in Fig.~\ref{fig:sim3snap}b).}
  \label{fig:speg}
\end{figure}

\section{Method} \label{sec:mtd}
This work aims to investigate the synthetic \CaIR\ Stokes profile of simulated chromospheric magnetic reconnection from realistic 2D radiative MHD simulations. The data are processed in two independent steps: First, we perform 2D RMHD simulations with RAMENS code \citep{Iijima2016} to reproduce the chromosphere in an active region. Then, we perform Stokes profile synthesis based on the simulated atmosphere and magnetic field with 1D RH code \citep{uiten2001}. 

\begin{figure*}[!t]
  \centering
  \includegraphics[width=\textwidth]{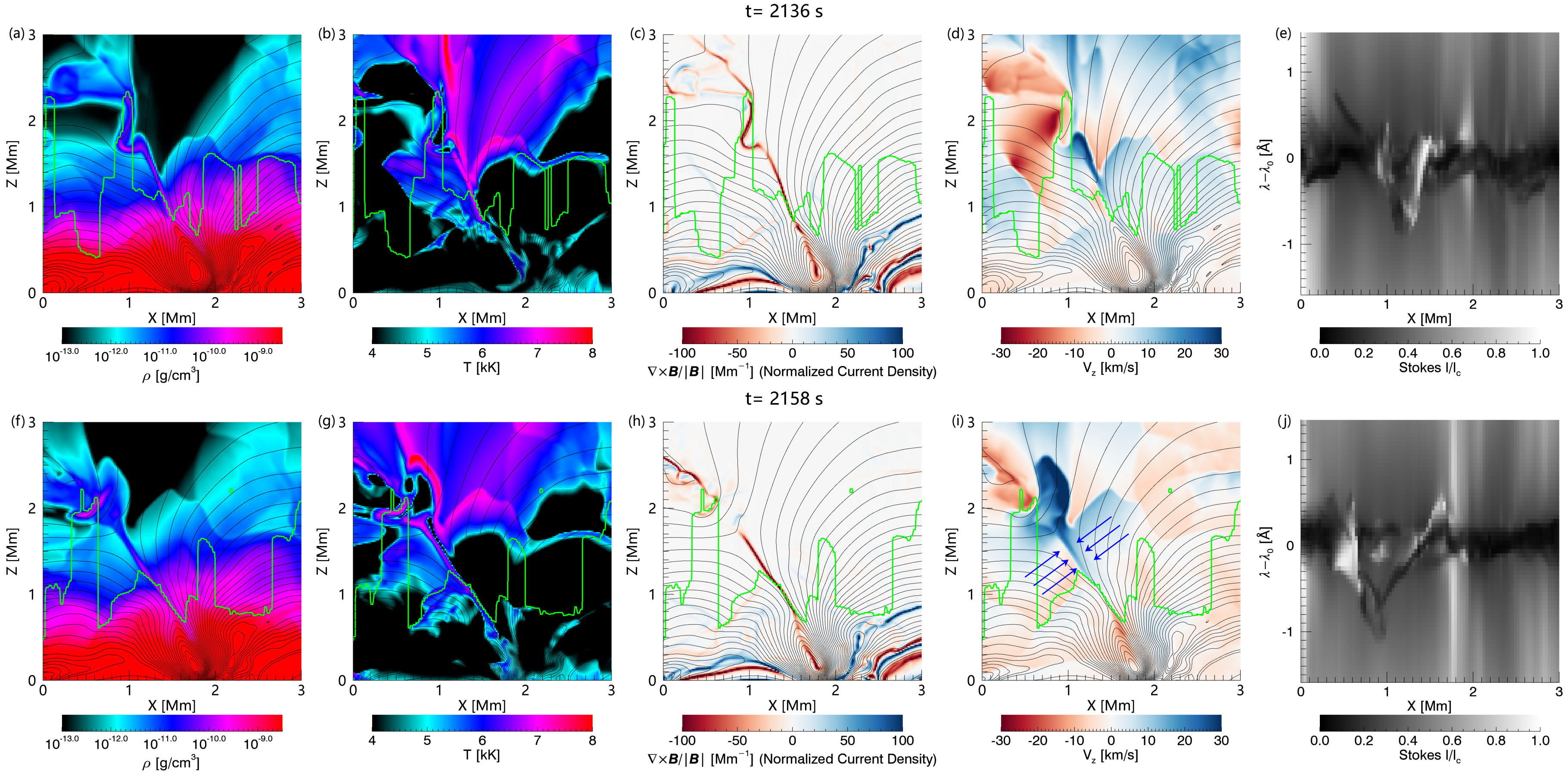}
  \caption{2D maps around the reconnection event 1 at $t=2136$~s and $t=2158$~s for (a)(f) density, (b)(g) temperature, (c)(h) normalized current density, and (d)(i) $V_z$ along with (e)(j) synthetic Stokes~$I$. The location in the whole simulation domain is indicated by the yellow box in Fig.~\ref{fig:sim3snap}c. Blue arrows in panel (i) indicate the velocity jump as the signal of shocks. A movie with time evolution from $t=1800$~s to $t=2300$~s is available online. The upward outflow appears 3 times when the Petschek-like reconnection is triggered by the passing shocks, at $t=1870\text{--}2000$~s, $t=2120\text{--}2200$~s, and $t=2260\text{--}2300$~s. However, the downward outflow is negligible except for the second time. After that the bipolar field topology disappears.}
  \label{fig:mrx1:viewI}
\end{figure*}
\subsection{Simulation} \label{sec:mtd:sim}
The simulation is implemented with RAMENS code \citep{Iijima2016}, which includes realistic convection motion, anisotropic heat conduction, optical-thick radiative transfer below the photosphere, optical-thin radiative loss model above the chromosphere \citep{Goodman12,Landi12}, and equation of state assuming local thermodynamic equilibrium (LTE) with ionization. The detailed description of this code is explained in \citet{Iijima2015}. 

The simulation box is set as a two-dimensional rectangle in $xz$-plane with $10$~Mm horizontal ($x$) range and $16$~Mm vertical ($z$) range. The simulation domain covers the convection zone from $z=-2$~Mm, the photosphere, the chromosphere and the corona up to $z=14$~Mm, with $z=0$~Mm set as the initial solar surface. The grid spacing is set to uniformly $20$~km, approximately. A hot plate of $2$~MK is placed at the upper boundary to maintain the temperature of the corona. 

To mimic an emerging flux in an active region, we proceed the simulation with two steps. In the first step, the simulation domain is initialized with uniform vertical background magnetic field ($B_{\mathrm{bkg}}>3$~G) (Fig.~\ref{fig:sim3snap}a). The simulation runs for $11000$~s to reach the well-relaxed state, then as the second step, horizontal magnetic flux sheet with uniform magnetic strength $B_{\mathrm{Flux}}>100$~G is imposed in $-2$~Mm $<z<-1$~Mm in the convection zone (Fig.~\ref{fig:sim3snap}b), at which the timestamp is set as $t=0$~s. After that the simulation continues to run for the other $4000$~s, during which the magnetic flux reaches the chromosphere due to magnetic buoyancy and leads to magnetic reconnection eventually (Fig.~\ref{fig:sim3snap}c).

We implemented various combinations of $B_{\mathrm{bkg}}$ and $B_{\mathrm{Flux}}$ to reproduce multiple reconnection events with different field topology. However, the results did not reveal any notable differences that align with our interests, except for the existence of plasmoids. Among the 7 reconnection events in 4 simulations, plasmoids appear in 4 events, while the other 3 events do not exhibit them.

To indicate the most sensitive layer of the \CaIR\ line as \citet{delaCR17}, we present lime contour lines representing the $\tau=1$ surface based on the mean optical depth in the wavelength range $\lambda\sim\lambda_0\pm0.25$~\AA. These contour lines are displayed in all figures of the simulation data in the remainder of this paper. It should be noted that the height of this surface varies significantly across horizontal locations, owing to the line of sight (LOS) velocity-induced Doppler shift of the surrounding plasmas, which may frequently exceed the specified wavelength range. For Stokes features outside this wavelength range, the formation height should be inferred from the LOS velocity map, not from these contour lines. 

\subsection{Stokes Profile Synthesis} \label{sec:mtd:syn}
The Stokes profile synthesis is implemented with 1D RH code \citep{uiten2001} which solves radiation transfer equations with general non-LTE conditions. The Calcium atom model is considered with an assumption of complete frequency redistribution (CRD) \citep{uiten1989}. The line-of-sight is set to vertical negative $z$ direction to mimic the solar disk center observation. Polarization signal is exclusively induced by Zeeman effect. Scattering polarization is not included so as to increase the probability of iteration convergence. Fig.~\ref{fig:speg} shows an example of Stokes profile synthesized from the snapshot at $t=2$~s (Fig.~\ref{fig:sim3snap}b). In the remainder of this paper, the central wavelength on a static profile of \CaIR\ is defined as $\lambda_0=$8542.09~\AA, the continuum intensity $I_c$ is defined as the average intensity in the range of $8460$~\AA\ $<\lambda<8490$~\AA\ and $8675$~\AA\ $<\lambda<8700$~\AA, in which the intensity is nearly constant for each profile.

Beyond the default setup of the RH code, we computed the synthetic spectral profile on additional 3001 wavelength points on each simulation snapshot, spanning $8540.5$~\AA\ to $8543.5$~\AA\ with $1$~m\AA\ spacing, and then performed Gaussian smoothing with $50$~m\AA\ spacing to mimic the instrumental spectral resolution (See Section~\ref{sec:dis:dkist}). In addition, the input simulation data is smoothed with doubled grid size to suppress numerical noises in the simulation data, resulting in a final horizontal spatial smapling of $\sim40$~km.


\section{Result} \label{sec:res}
In a typical simulation run, the magnetic flux gradually emerges from the convection zone into the photosphere after several hundred seconds. The imposed magnetic energy then initiated the rapid expanding motion of chromospheric magnetic bubbles (e.g., the magnetic structure around $2$~Mm $<x<6$~Mm and $0$~Mm $<z<5$~Mm in Fig.~\ref{fig:sim3snap}c). Reconnections occur when the emerging magnetic flux encounters the pre-existing background field, or the other part of emerging flux. Typically, each chromospheric reconnection event lasts for $\sim300$~s and lead to upward outflows with velocities $>30$~km~s$^{-1}$. However, the downward outflow is usually suppressed by the dense lower chromosphere, leading to a lower outflow speed $\sim20$~km~s$^{-1}$. Moreover, magnetic islands can block outflows, restricting their spatial scale in Stokes signal. Instead, Stokes features of magnetic islands appear near the signal of outflow. Here we describe two typical reconnection events separately: one Petschek-like reconnection without magnetic islands, and one plasmoid-like reconnection with magnetic islands, to illustrate the characteristic Stokes features of chromospheric reconnection events.

\subsection{Reconnection Event 1: Petschek-like Reconnection Without Magnetic Islands} \label{sec:res:rxe1}

\begin{figure*}[!t]
  \centering
  \includegraphics[width=\textwidth]{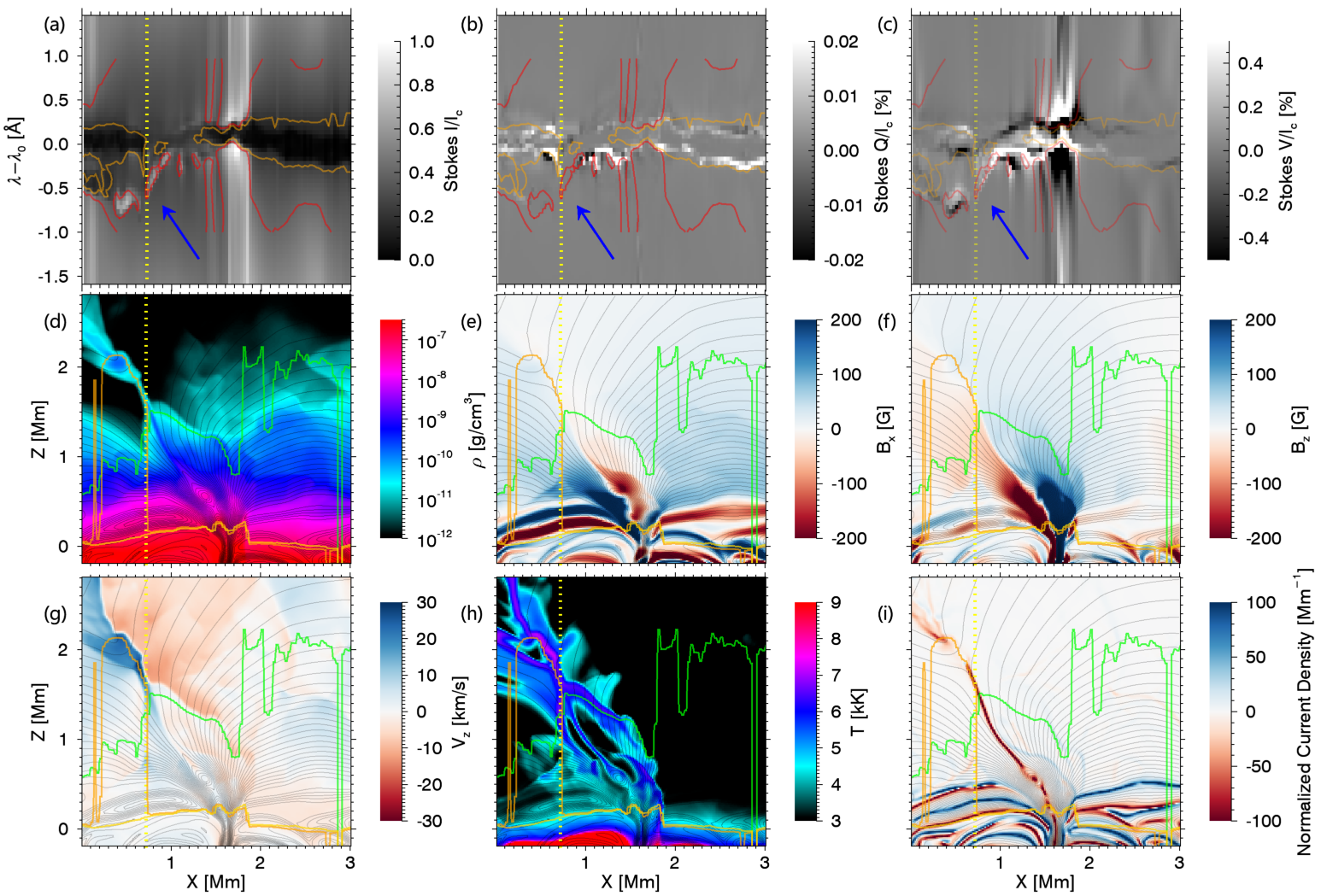}
  \caption{$x-\lambda$ maps for (a) Stokes~$I$, (b) $Q$, (c) $V$ with 2D maps for simulation data on (d) density, (e) $B_x$, (f) $B_z$, (g) $V_z$, (h) temperature, and (i) normalized current density around the reconnection event 1 at $t=1966$~s. The colored contour lines of Stokes~$I$ at $I=0.2I_c$ (orange) for absorption features and $I=0.4I_c$ (red) for emission features are shown on every map for Stokes signals. The blue arrows on panel (a), (b) and (c) indicate the Stokes features from the reconnection current sheet. The vertical yellow dotted line on each panel indicates the location of Fig.~\ref{fig:mrx1:iq-t}. Besides the lime contour line of the mean optical depth $\tau=1$ surface in the spectrum core on the simulation data panels, orange and gold contour lines are added for wavelength range $\lambda_0-0.5$~\AA$\sim-1.0$~\AA, and $\lambda_0+0.5$~\AA$\sim+1.0$~\AA, respectively.}
  \label{fig:mrx1:iqv}
\end{figure*}

In a simulation with $B_{\mathrm{bkg}}=3$~G and $B_{\mathrm{Flux}}=100$~G, this reconnection event begins around $t=1830$~s, when the emerging flux around $x\sim1$~Mm interacts with the other part of the emerging flux around $x\sim3$~Mm, resulting in a current sheet with a length of $\sim1$ Mm and an inclination of $\sim20^\circ$ (e.g., Fig.~\ref{fig:mrx1:viewI}c\&h). Here the normalized current density is defined as ${\boldsymbol{\nabla\times}\mathbf{B}}/{|\mathbf{B}|}$, similar to the definition of the "inverse characteristic length of the magnetic field" in \citet{NS2016}, which could indicate the location of the current sheet. Two slow shocks appear as density, temperature and velocity jump (e.g., Fig.~\ref{fig:mrx1:viewI}i) on both sides of the current sheet, indicating a Petschek-type reconnection \citep{Petschek1964}. (Note, however, the current sheets which should be co-located at the shocks are not well divided in Fig.~\ref{fig:mrx1:iqv}i, probably due to the closeness to the X-point and a lack of numerical resolution.) 

The reconnection outflow is continually ejected outward from $z\sim1.0$~Mm along the reconnection current sheet with peak speed $\sim40$~km~s$^{-1}$ around $t=2158$~s (Fig.~\ref{fig:mrx1:viewI}i). The downward reconnection outflow is suppressed by the dense photosphere, thus the speed of downward outflow only reaches $\sim20$~km~s$^{-1}$. During the reconnection process, the temperature of the plasma around the reconnection current sheet reaches up to $>7$~kK around $t=2136$~s (Fig.~\ref{fig:mrx1:viewI}b). The outflow dense plasma fades twice before the anti-parallel magnetic geometry fully disappears when $t>2300$~s, one around $t=2050$~s, the other around $t=2222$~s, which indicates the reconnection process pauses due to the intermittent emergence of magnetic flux. Counting only the time that outflow dense plasma exists, the timescale of this reconnection event is $\sim300$~s, which is much shorter than estimated reconnection time $t_R={L^2}/({V_A\delta})>700$~s from the Sweet-Parker model \citep{Parker1957}, in which $L\sim1.5$~Mm is the length of reconnection current sheet, $V_A\sim40$~km~s$^{-1}$ is the local Alfv\'en speed, and $\delta\sim80$~km is the thickness of reconnection current sheet. This gap on reconnection time indicates a fast reconnection event.
\subsubsection{Emission and Doppler Shift in \texorpdfstring{Stokes~$I$}{Stokes~I}} \label{sec:res:rxe1:stokesi}

\begin{figure}
  \centering
  \includegraphics[width=9.0cm]{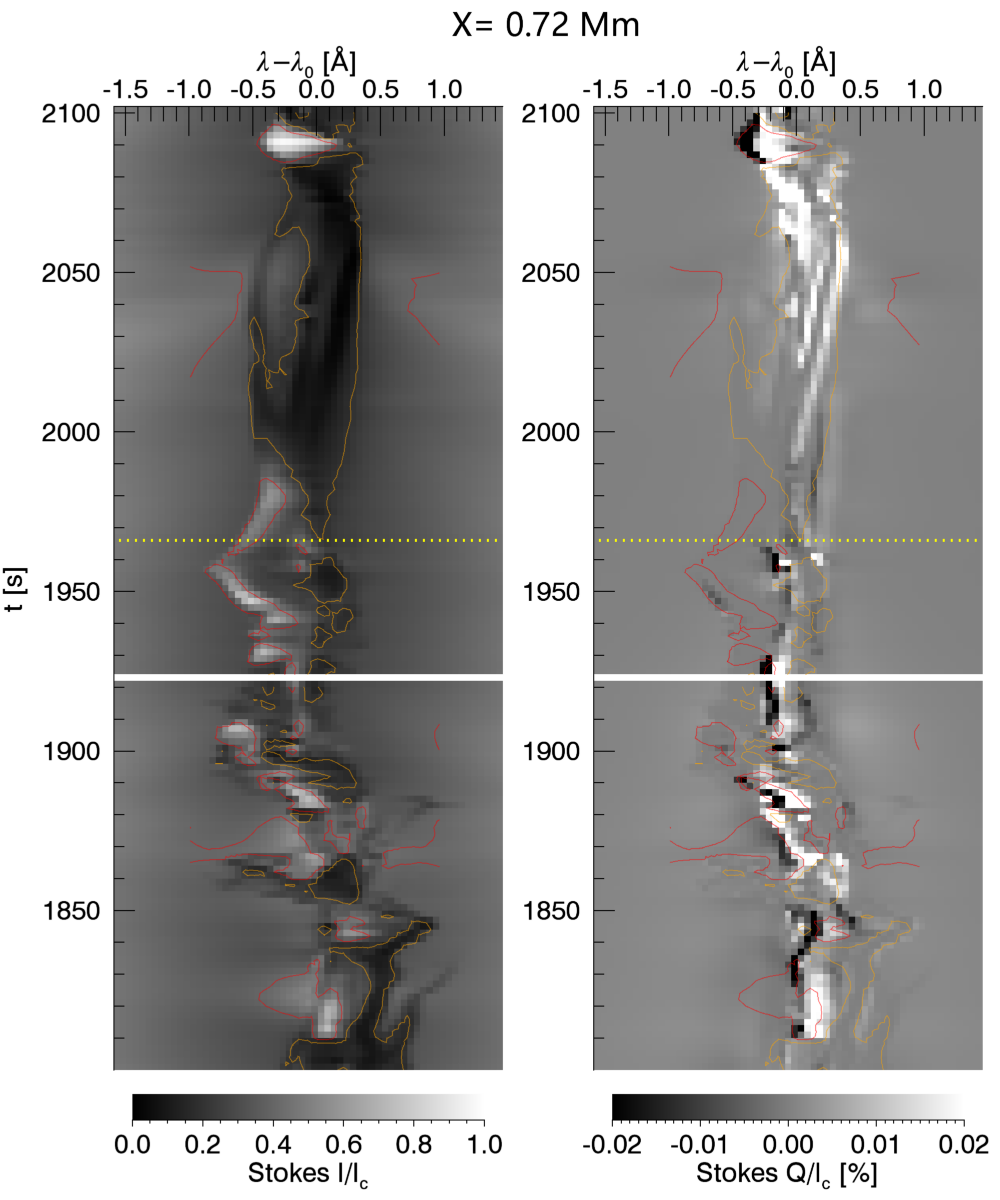}
  \caption{A $\lambda$$-t$ map for Stokes~$I$ and $Q$ on $x=0.72$~Mm. The colored contour line is the same as Fig.~\ref{fig:mrx1:iqv}. The 'white' horizontal line at $t=1922$~s represents missing data due to a failure in Stokes profile synthesis. The horizontal yellow dotted line on each panel indicates the time of Fig.~\ref{fig:mrx1:iqv}.}
  \label{fig:mrx1:iq-t}
\end{figure}

Here we describe several synthetic Stokes~$I$ features related to this reconnection event. Emission appears on the Stokes~$I$ at $t=2136$~s (Fig.~\ref{fig:mrx1:viewI}e) from $x=1.2$~Mm to $x=1.6$~Mm as the maximum temperature of corresponding reconnection sheet rises to $>7$~kK (Fig.~\ref{fig:mrx1:viewI}b). Here the temperature has reached its peak while the reconnection outflow is still developing. Later, the Stokes~$I$ at $t=2158$~s (Fig.~\ref{fig:mrx1:viewI}j) shows $\Delta\lambda\sim-1.1$~\AA\ blue-shifted features on $x\sim0.9$~Mm and $\Delta\lambda\sim+0.6$~\AA\ red-shifted features on $x\sim1.7$~Mm (relative wavelength $\Delta\lambda$ is defined as $\Delta\lambda=\lambda-\lambda_0$ in the remainder of this paper), as the simulation data shows bidirectional reconnection outflow on the same spatial horizontal location (Fig.~\ref{fig:mrx1:viewI}i). In detail, the upward reconnection outflow reaches speed of $\sim40$~km~s$^{-1}$ at $(x,z)\sim(0.9,1.8)$~Mm, resulting in the $\Delta\lambda\sim-1.1$~\AA\ blue-shifted absorption feature. The downward reconnection outflow is suppressed by the dense plasma below, thus only reaches speed of $\sim20$~km~s$^{-1}$ at $(x,z)\sim(1.7,0.6)$~Mm, therefore the corresponding red-shifted emission is only up to $\Delta\lambda\sim+0.6$~\AA. This asymmetry in the Doppler shift of outflow is also reported in observations such as \citet{Matsumoto08} and \citet{Vissers19}. In addition, the Stokes~$I$ (Fig.~\ref{fig:mrx1:viewI}j) shows an inclined line-shape feature from $x=0.9$~Mm to $x=1.6$~Mm, which indicates the outflow accelerating process inside the reconnection current sheet. From $x=0.4$~Mm to $x=0.9$~Mm, the inclined line-shape feature indicates the deaccelerating process of upward outflow after ejected out of the reconnection current sheet. 

\subsubsection{Sign Reversal in \texorpdfstring{Stokes~$V$}{Stokes~V}} \label{sec:res:rxe1:stokesv}

The Stokes~$V$ profile of a magnetized static atmosphere shows two opposite peaks in the red and blue side of the core \cite[e.g.,][]{delTI03,Pietarila06,delaCR13,QN2016}. The sign of positive-negative peak depends on the polarity of longitudinal magnetic field \citep[Fig.~\ref{fig:speg};][]{Landi04}. In this work, the LOS of synthesis process is set along the negative $z$ direction, thus the sign of positive-negative peak depends on the sign of vertical magnetic field $B_z$. In the case of core absorption profile on Stokes I, when $B_z>0$, the Stokes~$V$ profile shows positive peak on blue side and negative peak on the red side, vice versa. Note that this behavior could be reversed if the Stokes~$I$ absorption feature turns into emission. If the sign of Stokes~$V$ profile is reversed while the Stokes~$I$ profiles remain in absorption (or emission), one can infer the reversal of the vertical field polarity.

Here we represent this well-known mechanism as the Stokes~$V$ sign reversal features in this reconnection event. In Fig.~\ref{fig:mrx1:iqv}c, the Stokes~$V$ at $t=1966$~s from $x=0.4$~Mm to $x=0.7$~Mm shows positive peak at $\Delta\lambda\sim+0.1$~\AA\ and negative peak at $\Delta\lambda\sim-0.1$~\AA\ around the Stokes~$I$ (Fig.~\ref{fig:mrx1:iqv}a) absorption feature at $\Delta\lambda\sim0.0$~\AA, indicates that the corresponding vertical magnetic field $B_z<0$. To the right of the reconnection sheet from $x=0.7$~Mm to $x=1.1$~Mm, as the Stokes~$I$ line core remain in absorption, the Stokes~$V$ shows negative peak at $\Delta\lambda\sim+0.1$~\AA\ and positive peak at $\Delta\lambda\sim-0.2$~\AA, indicates that the local vertical magnetic field $B_z>0$. For the appearance, we see white-black features switching to black-white feature, as the sign reversal feature. Since this feature is around the line core, the formation height could be indicated by the lime contour line of optical depth on each panel. As this line crosses the current sheet, the polarity of the vertical magnetic field $B_z$ (Fig.~\ref{fig:mrx1:iqv}f) on the lime contour line is reversed at the cross point $x=0.75$~Mm, resulting in the sign reversal feature on Stokes~$V$ at the same horizontal location. As the Stokes~$I$ profiles remain in absorption, this Stokes~$V$ sign reversal feature indicates that the vertical magnetic field reverses its polarity at the spectral line formation height, which usually forms the field topology for magnetic reconnection. 

In contrast, regarding the outflow accelerating process inside the current sheet (see Section~\ref{sec:res:rxe1:stokesi}), signals from current sheet are usually highly Doppler shifted (e.g. $(x,\Delta\lambda)\sim(0.5$~Mm,$-0.7$~\AA$)$ in Fig.~\ref{fig:mrx1:iqv}a), which are far from the sign reversal feature around $\Delta\lambda\sim0.0$~\AA. Thus the Stokes~$V$ sign reversal features here are not related with the plasma inside the current sheet. 

\subsubsection{Stokes~$Q$/Linear Polarization (LP) Reduction} \label{sec:res:rxe1:lpreduction}

The Stokes~$Q$/$U$ profile of a magnetized static atmosphere shows two peaks in the same sign in both red side and blue side of the core (see Fig.\ref{fig:speg} for example). The amplitude of Stokes~$Q$/$U$ peaks depends on the strength of transverse magnetic field \citep{ctno2018}. In this work, based on Zeeman effect, Stokes~$Q$ signal is mainly from $B_x$, while Stokes~$U$ signal is mainly from $B_y$. As $B_y$ is set as $0$ in our 2D simulation, Stokes $U$ is dominated by magneto-optical effects \citep{delTI03}, which is highly correlated to Stokes~$Q$. Thus, in this work we only discuss Stokes~$Q$ for the linear polarization. 

In Fig.~\ref{fig:mrx1:iqv}a, the map for the Stokes~$I$ at $t=1966$~s from $x=0.4$~Mm to $x=1.0$~Mm shows an inclined emission feature, which is indicated by the red contour line and the blue arrow on the figure. Based on the $V_z$ data (Fig.~\ref{fig:mrx1:iqv}g), it could be inferred that this feature comes from the reconnection current sheet. At the same coordinate marked by the red contour line, the Stokes~$Q$ signal (Fig.~\ref{fig:mrx1:iqv}b) related to reconnection current sheet is noticeably weaker than surrounding region (e.g., the Stokes~$Q$ signal around $\Delta\lambda\sim0.0$~\AA\ marked by the yellow contour line). These Stokes~$Q$ signals are in the same order of $Q$ signals in the far wing (e.g., $|\Delta\lambda|\sim1.5$~\AA), which is only $1$\% of $Q$ signals from the background atmosphere (e.g., $|\Delta\lambda|\sim0.1$~\AA\ in Fig.~\ref{fig:mrx1:iqv}b). This feature is named as Stokes~$Q$ reduction in this work, which is coincident with the Stokes~$I$ feature of reconnection current sheet. Stokes~$Q$ reduction feature indicates that the local horizontal magnetic field is relatively weaker than surrounding region, which is consistent with the weak horizontal magnetic field $B_{\mathrm{h}}=\sqrt{{B_x}^2+{B_y}^2}<10$~G (as $B_y=0$ in this work) (Fig.~\ref{fig:mrx1:iqv}e) in the center of the current sheet. 

Fig.~\ref{fig:mrx1:iq-t} shows time evolution of the Stokes~$Q$ reduction on $x=0.72$~Mm. From $t=1930$~s to $t=1990$~s, the emission Stokes~$I$ feature on the blue side leads to weak signal on Stokes~$Q$, regarded as Stokes~$Q$ reduction as above. Also, one could recognize the absorption Stokes~$I$ feature around $\Delta\lambda\sim-0.4$~\AA\ from $t=1990$~s to $t=2070$~s, this Stokes feature also coexists with Stokes~$Q$ reduction, indicating the reconnection current sheet. In addition, some Stokes~$Q$ signals appear around $\Delta\lambda\sim-0.6$~\AA\ at $t\sim1950$~s, which occurs when the upper end of the reconnection current sheet reaches $x=0.72$~Mm in the developing state.

We proposed the following explanation for Stokes~$Q$ reduction: the reconnection process drains plasmas from surrounding region and concentrates them into the center of the current sheet. These plasmas occupy only a thin layer on spatial height but are optically thick, thus the physical condition of this thin layer could significantly affect the Stokes profile. Due to the weak transverse field inside the reconnection current sheet, the LP signal contributions are minimal, resulting in the Stokes~$Q$/$U$ reduction feature. This reduction feature could indicate the existence of an optical thick region with weak transverse field, which usually appears in the center of reconnection current sheet. 

Note that a similar reduction feature is not seen in the Stokes~$V$ map (Fig.\ref{fig:mrx1:iqv}c), as in the weak field regime, Stokes~$V$ decreases more slowly than Stokes~$Q$/$U$ with decreasing transverse field \citep{ctno2018}.

In general cases of anti-parallel magnetic reconnections, Stokes~$U$ should show similar behavior as Stokes~$Q$ reduction when all the magnetic components are annihilated inside the current sheet, therefore Stokes~$Q$ reduction could be summarized as LP reduction. We suggest that in reconnection events with 2D field geometry, LP reduction could be one possible Stokes feature indicating the negligible magnetic field inside the reconnection current sheets, in the absence of plasmoids.

\begin{figure}[t]
  \centering
  \includegraphics[width=9.0cm]{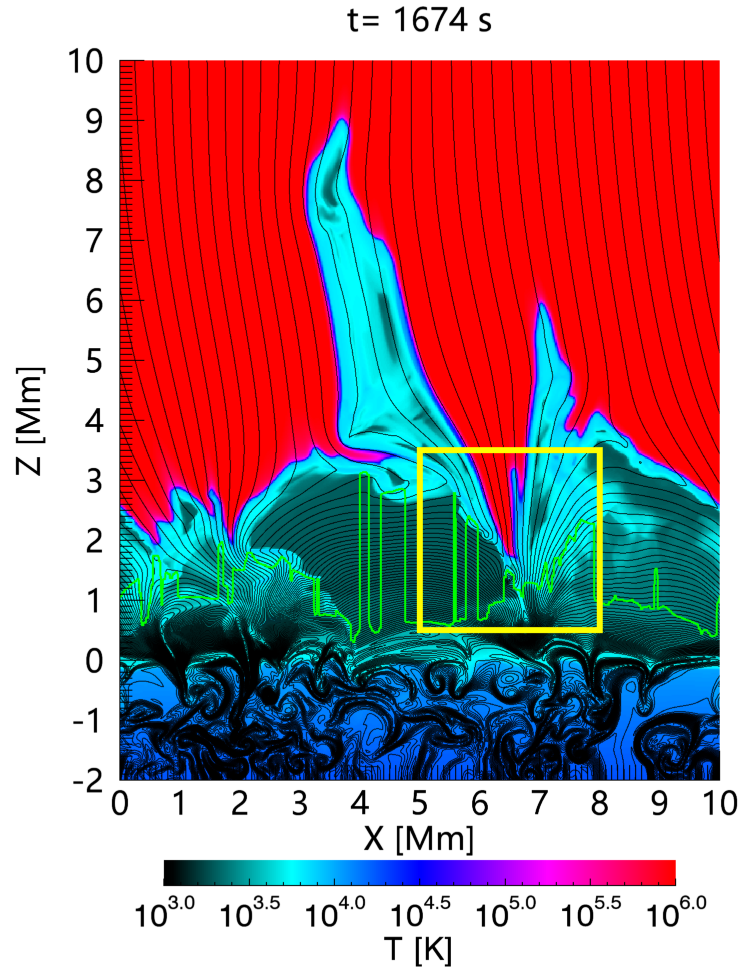}
  \caption{Temperature of the whole simulation domain at $t=1674$~s around reconnection event 2. The yellow box indicates the location of 2D maps in Fig.~\ref{fig:mrx2:viewI}.}
  \label{fig:mrx2:snap}
\end{figure}

\begin{figure*}[t]
  \centering
  \includegraphics[width=\textwidth]{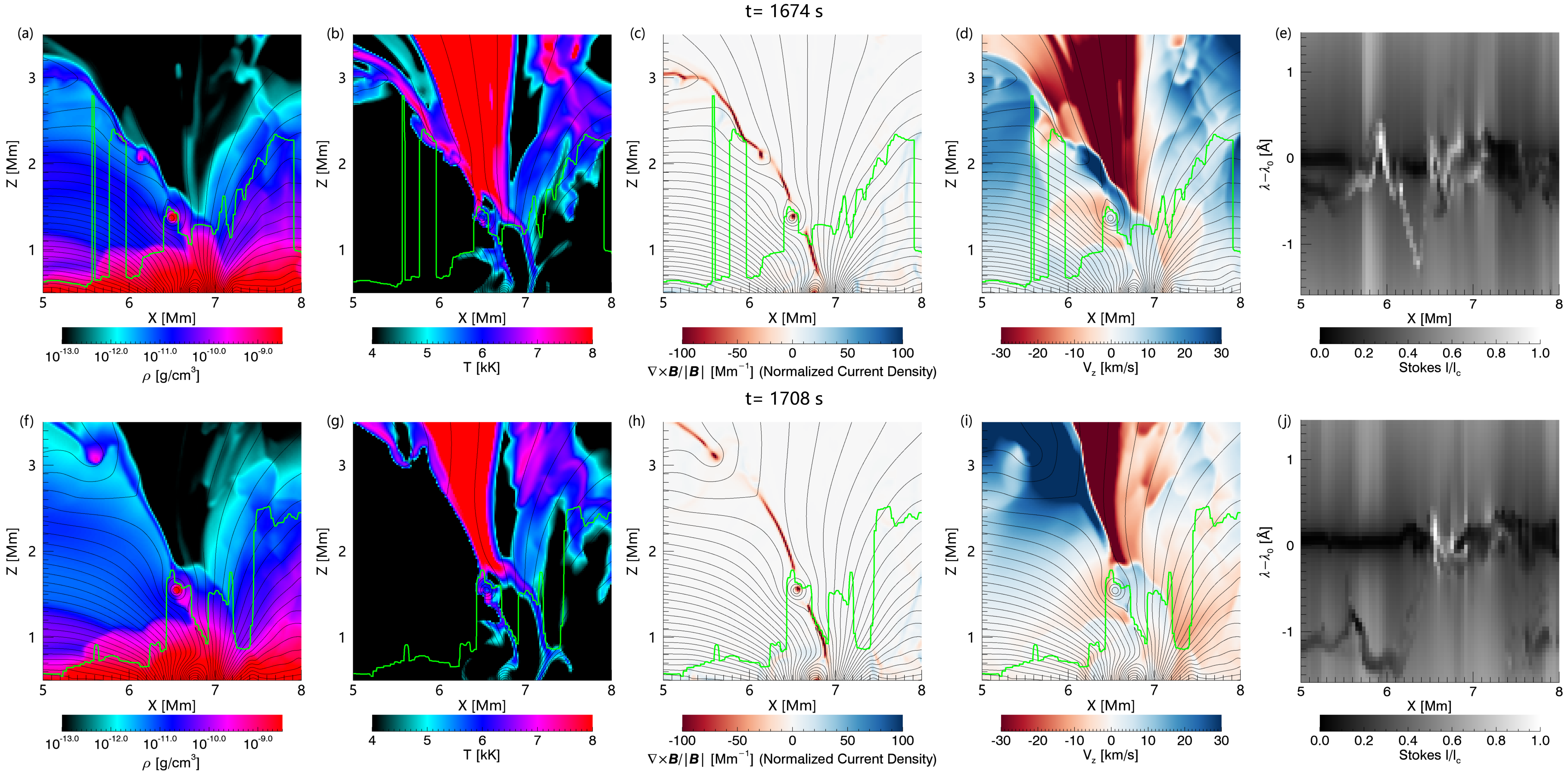}
  \caption{2D maps around the reconnection event 2 at $t=1674$~s and $t=1708$~s organized similar to Fig.~\ref{fig:mrx1:viewI}. The location in the whole simulation domain is indicated by the yellow box in Fig.~\ref{fig:mrx2:snap}. A movie with time evolution from $t=1600$~s to $t=2000$~s is available online. The bidirectional outflow appears from $t=1620$~s. Then a huge magnetic island forms by $t=1660$~s, blocking the downward outflow. At $t=1800$~s, the passing shocks push down the magnetic island and trigger the latter part of the reconnection, leading to several tiny blobs ejecting upward. The bipolar field topology disappears after $t=1930$~s.}
  \label{fig:mrx2:viewI}
\end{figure*}

\begin{figure*}[t]
  \centering
  \includegraphics[width=\textwidth]{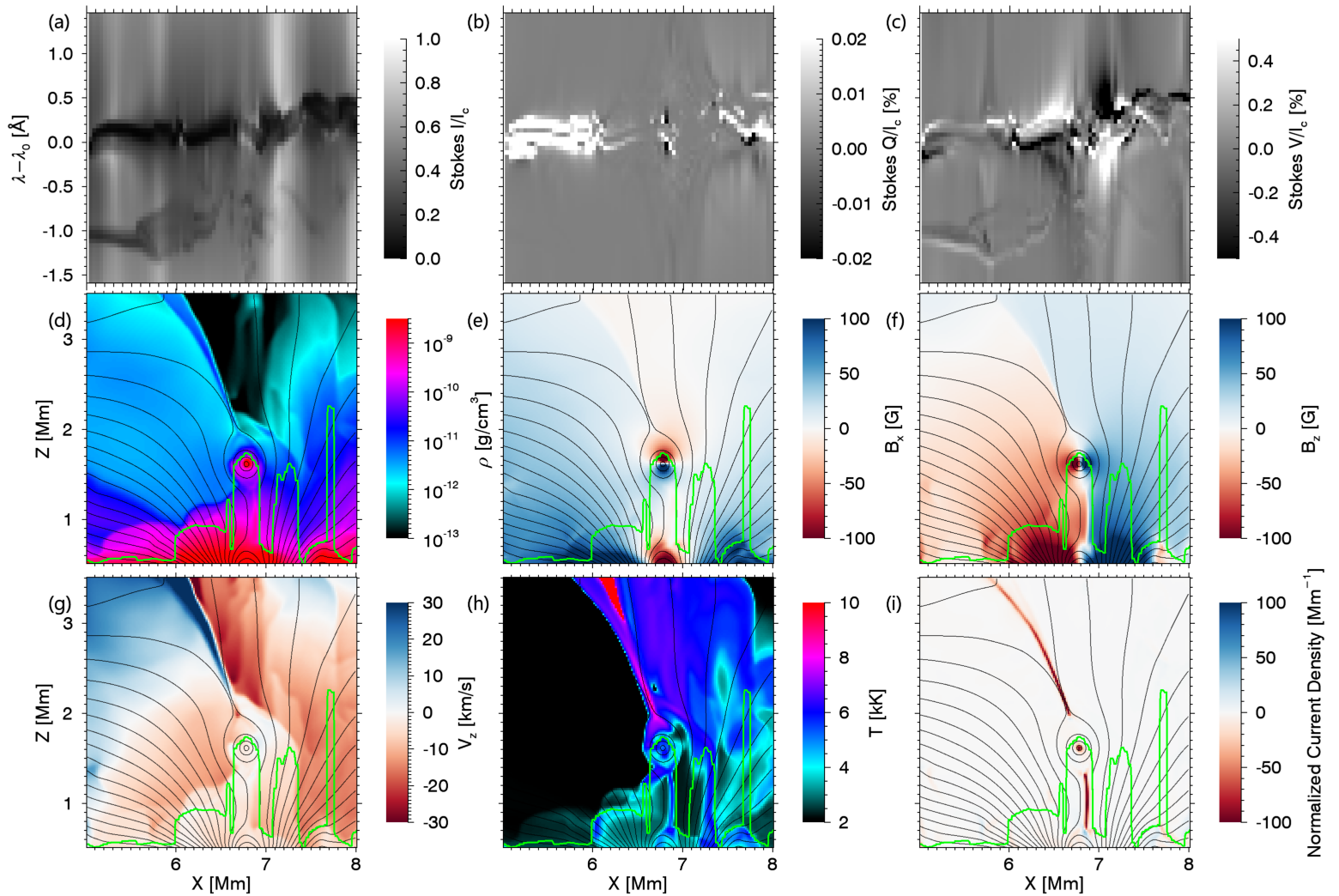}
  \caption{Stokes signals and simulation data around reconnection event 2 at $t=1748$~s organized similar to Fig.~\ref{fig:mrx1:iqv}.}
  \label{fig:mrx2:iqvo}
\end{figure*}

\begin{figure*}[!t]
  \centering
  \includegraphics[width=\textwidth]{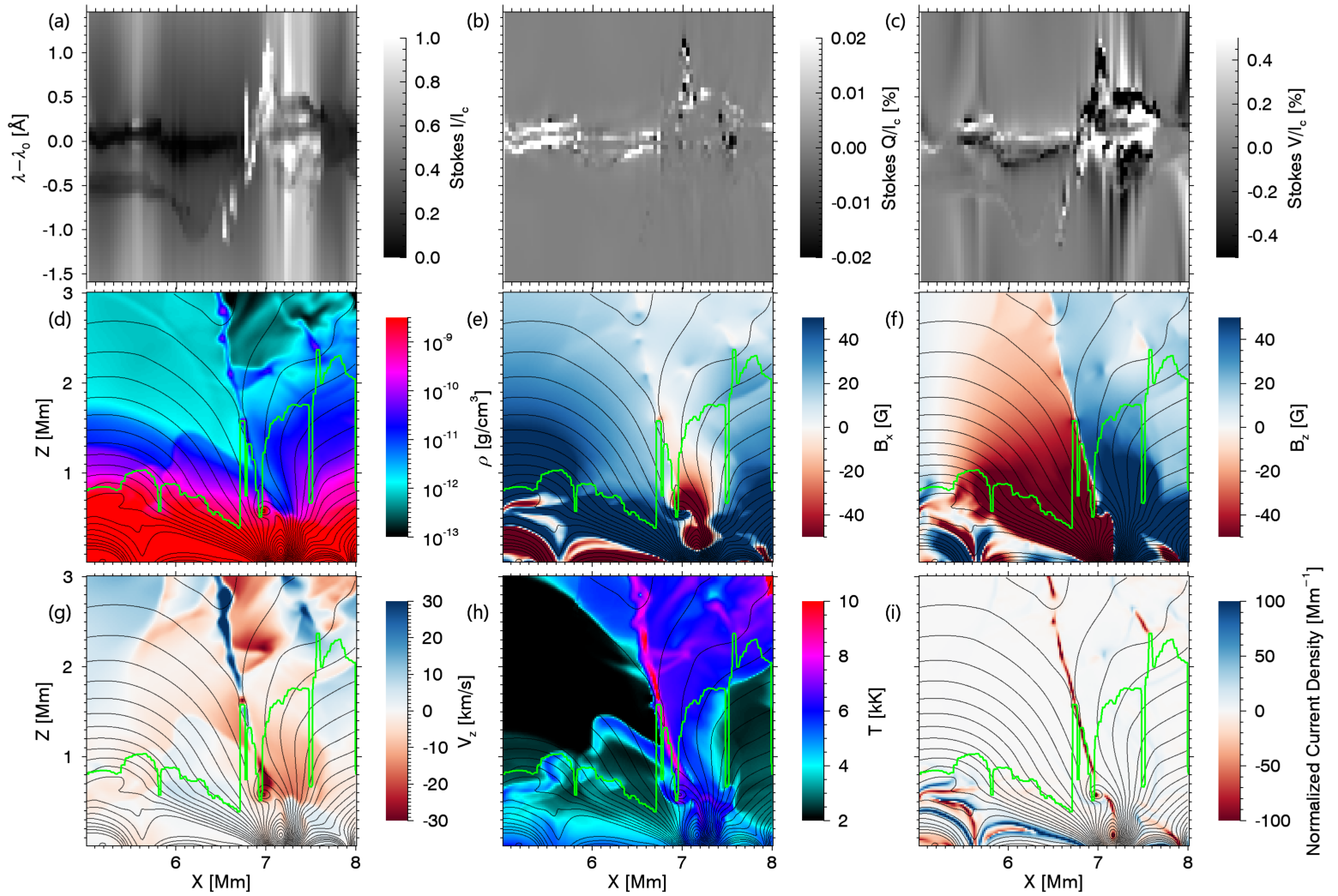}
  \caption{Stokes signals and simulation data around reconnection event 2 at $t=1826$~s organized similar to Fig.~\ref{fig:mrx1:iqv}.}
  \label{fig:mrx2:soiqv}
\end{figure*}

\subsubsection{Wing Enhancements and Sign Reversals in Photospheric Reconnection} \label{sec:res:rxe1:wingsig}

Extreme signals appear on the wing part of Stokes~$I$ and $V$ from $x=1.6$~Mm to $x=1.9$~Mm (Fig.~\ref{fig:mrx1:iqv}a\&c), including wing enhancements similar to Ellerman bombs and some brightening near the absorption core on the Stokes~$I$ map. The formation height of the wing enhancement could be indicated by the orange and gold contour lines, which represent the mean optical depth surface on the wing. These enhancements are caused by heating from the wandering reconnection current sheet around $z\sim0.5$~Mm, which raises the temperature to $T\sim6$~kK from $x=1.6$~Mm to $x=1.9$~Mm on the formation height (Fig.~\ref{fig:mrx1:iqv}h). Moreover, an extra passing shock heats the plasma around $x=1.8$~Mm to $T\sim7$~kK, causing even stronger signals. Note that a plasmoid appears around $x=1.5$~Mm (Fig.~\ref{fig:mrx1:iqv}i) on the formation height, but its temperature does not exceed $T\sim5$~kK, thus show no recognizable signals. The velocities show predominantly downward motions in the surrounding region, except around the plasmoid, where locates the upward outflow from the current sheet $z\sim0.0$~Mm. The amplitude is no more than $5$~km~s$^{-1}$. As for the Stokes~$V$ profile, the wing signal shows two sign reversals, corresponding to the field polarity concentrations in the photosphere. 

The brightening near the absorption core forms higher than the wing part but lower than the absorption core. These signals are from the high-temperature region around $(x,z)\sim(1.7,0.7)$~Mm (Fig.~\ref{fig:mrx1:iqv}h), which is first pre-heated by the wandering reconnection current sheet, and then heated by the passing shocks mentioned above. Meanwhile, the strong Stokes~$V$ signal indicates a unipolar magnetic field. These signals can be considered as a consequence of magnetic reconnection slightly above the photosphere, but not directly related to the current sheet. 

In summary, these extreme signals on the wing arise from the photospheric part of the reconnection, with additional heating provided by a passing shock. However, due to the presence of multiple field polarity reversals in the surrounding regions, it could be challenging to precisely locate the current sheet from Stokes maps. In addition, the brightening near the absorption core indicates heating slightly above the photosphere, but lower than the general chromosphere, resulting in double-peaked emissions near the core on Stokes~$I$ profile.

\subsection{Reconnection Event 2: Reconnection with Magnetic Islands} \label{sec:res:rxe2}

Fig.~\ref{fig:mrx2:snap} shows the general view of simulation domain in which reconnection event 2 appears, with $B_{\mathrm{bkg}}=10$~G and $B_{\mathrm{Flux}}=180$~G. This reconnection event occurs from $t=1612$~s when the emerging flux $x\sim6$~Mm interacts with pre-existing background magnetic field $x\sim7.5$~Mm. During the reconnection process, several magnetic islands appear. Especially, a huge magnetic island ($(x,z)=(6.5,1.5)$~Mm in Fig.~\ref{fig:mrx2:viewI}a\&f) is formed inside the current sheet (Fig.~\ref{fig:mrx2:viewI}c\&h), thus the reconnection current sheet is divided into two parts. The lower part ($(x,z)\sim(6.8,1.1)$~Mm in Fig.~\ref{fig:mrx2:viewI}h\&i) dissipates rapidly due to the shorter current sheet length with lower outflow speed and shorter spatial scale, while the higher part maintains a reconnection current sheet with longer lifetime. Here we focus on the current sheet located at an altitude higher than the magnetic island. 

In the early stage of the reconnection at $t=1708$~s (Fig.~\ref{fig:mrx2:viewI}i), upward reconnection outflow appears from $x=6.1$~Mm to $x=6.5$~Mm, reaching a maximum velocity of $\sim50$~km~s$^{-1}$. As a result, the Stokes~$I$ (Fig.~\ref{fig:mrx2:viewI}j) shows an inclined blue shifted absorption feature up to $\Delta\lambda\sim-1.4$~\AA\ from $x\sim6.1$~Mm to $x\sim6.5$~Mm. Meanwhile, similar blue shifted feature appears from $x\sim5$~Mm to $x\sim6$~Mm, which is from plasmas previously ejected out of the current sheet. The downward reconnection outflow ($(x,z)=(6.5,1.9)$~Mm in Fig.~\ref{fig:mrx2:viewI}i) is blocked by the huge magnetic island beneath, resulting in a narrow spatial scale $\sim0.1$~Mm but the outflow speed remains $\sim20$~km~s$^{-1}$. On the map for the Stokes~$I$ (Fig.~\ref{fig:mrx2:viewI}j), a narrow emission feature appears at $x=6.5$~Mm from $\Delta\lambda\sim-0.3$~\AA\ to $\Delta\lambda\sim+0.5$~\AA, corresponding to the blocked downward outflow. Between the blue-shifted absorption at $x=6.2$~Mm and the emission at $x\sim6.5$~Mm, the current sheet fades in the Stokes~$I$ signal. This indicates a decrease in Stokes~$I$ signal strength from the emission level to an intermediate level, suggesting a cooling process during the upward outflow ejection. 

Disregarding the signal around the huge magnetic islands, the polarization signals show similar behavior to Section~\ref{sec:res:rxe1:stokesv}\&\ref{sec:res:rxe1:lpreduction}. In Fig.~\ref{fig:mrx2:iqvo} at $t=1748$~s, a huge magnetic island is located from $x=6.6$~Mm to $x=7.0$~Mm. On the left side of this magnetic island, from $x=6.0$~Mm to $x=6.6$~Mm, the map for Stokes~$V$ (Fig.~\ref{fig:mrx2:iqvo}c) shows white-black features around $\Delta\lambda=+0.2$~\AA, while on the right side from $x=7.0$~Mm to $x=7.3$~Mm it shows black-white features around $\Delta\lambda=+0.2$~\AA. Similar to Section~\ref{sec:res:rxe1:stokesv}, this feature indicates the polarity of the local vertical magnetic field (Fig.~\ref{fig:mrx2:iqvo}f) in the background atmosphere is reversed across the current sheet.

Regarding that on the map for Stokes~$I$ (Fig.~\ref{fig:mrx2:iqvo}a), the inclined feature from $(x,\Delta\lambda)=(6.3$~Mm$,-1.0$~\AA$)$ to $(x,\Delta\lambda)=(6.7$~Mm$,0.0$~\AA$)$ is from the current sheet (Fig.~\ref{fig:mrx2:iqvo}i). Similar to Section~\ref{sec:res:rxe1:lpreduction}, Stokes~$Q$ (Fig.~\ref{fig:mrx2:iqvo}b) on the same coordinate shows unrecognized weak signal, except for signals around the magnetic islands. 

\subsubsection{Stokes Features of Magnetic Islands} \label{sec:res:rxe2:mgilad}

Several magnetic islands appear during the reconnection process, including huge ones ($\sim300$~km) and tiny ones ($\sim40$~km). The huge magnetic island $x\sim6.8$~Mm which exists from $t=1630$~s to $t=1826$~s moves slowly toward the photosphere during its lifetime. Regarding the strong magnetic field inside of it, both Stokes~$V$ and Stokes~$Q$ (Fig.~\ref{fig:mrx2:iqvo}b) show strong signal around this magnetic island. However, regarding that the simulation resolution in this study is too low to resolve the fine inner structure of magnetic islands \citep{ni2020}, the amplitude of polarization signals might be overestimated.

After the magnetic island moves into the dense photosphere, the upper current sheet (Fig.~\ref{fig:mrx2:soiqv}i) is elongated, then several tiny magnetic islands (Fig.~\ref{fig:mrx2:soiqv}d\&e\&f) appear in the current sheet at $t=1826$~s. These magnetic islands are accelerated to $>30$~km~s$^{-1}$ (Fig.~\ref{fig:mrx2:soiqv}g) and eject out of the current sheet while several corresponding fragmented emissions appear on the Stokes~$I$ (Fig.~\ref{fig:mrx2:soiqv}a), e.g., features at $x=6.55$~Mm, $x=6.65$~Mm, $x=6.8$~Mm. In addition, these Stokes~$I$ features show broad ($>0.4$~\AA) profiles similar to \citet{Innes15}. This broad emission also appears around the huge magnetic islands (Fig.~\ref{fig:mrx2:iqvo}a) but could hardly be seen since its huge mass suppresses the heating and accelerating by the magnetic reconnection. As for the polarization signal, considerable Stokes~$V$ signals also appear around these tiny islands but the Stokes~$Q$ signals are much weaker because the Stokes~$Q$ signals depend on the horizontal magnetic field in a quadratic relationship while Stokes~$V$ signals are proportional to the vertical magnetic field \citep{ctno2018}.

\section{Discussion} \label{sec:dis}

\subsection{Applicability of Current Results for Observations} \label{sec:dis:dkist}
Here we summarize the amplitude and timescale of signals in our result and discuss the visibility of these features in real observations. Stokes~$Q$ signal is commonly weaker than the Stokes~$V$ signal in this study. Typical amplitude of Stokes~$Q$ in this study is $2\times10^{-4}I_c$, which is slightly higher than the sensitivity of state-of-the-art spectro-polarimeter $\sim10^{-4}I_c$. Moreover, polarization effects other than Zeeman effect have not been taken into consideration in this study, which may enhance the linear polarization signal for magnetic field $B_{\mathrm{h}}<100$~G \citep{Stepan16}. Since the LP reduction features highly rely on the weak LP signals due to weak field, the contrast between the reduced LP signals and surrounding LP signals may be saturated, making it challenging to detect these features in weak field regime ($B_{\mathrm{h}}<100$~G). Still, the amplitude of Stokes~$Q$/$U$ signal from Zeeman effect scales with the square of transverse field ($Q\propto{B_{\mathrm{h}}}^2$), since the chromospheric magnetic field in our simulation is only $\sim100$~G, we could expect $>20$ times stronger signals in real observations when $B\sim300\text{--}500$~G in active regions \citep{Pietrow20,daSS22,Vissers22,Judge24}. This could enhance the visibility of general LP signals and strengthen the amplitude contrast necessary to capture LP reduction features. Therefore, in real solar observations, the Stokes features in this study could be observable in plage regions.

The spatial grid size of our simulation is $\sim20$ km, for the synthetic profile the spatial grid size is doubled as $\sim40$~km, which is close to the sampling of DKIST ViSP ($<50$~km) \citep{deWijn22}. \citet{Moe2022} concludes that to compare with observation result, the spatial resolution of numerical simulations should be at least twice the spatial sampling on Stokes signal to resolve the small-scale structures, thus our simulation is under an appropriate spatial resolution to compare with DKIST observations. In our study, we adopted an effective spectral resolution on which the minimum wavelength interval is $50$~m\AA, which is similar to that of DKIST ViSP: $\sim47$~m\AA\ on $8542$~\AA. Although we did not consider the time cadence in observation, the lifetime of features in this study is larger than $50$~s, which is much longer than the cadence of full Stokes profile observation in DKIST ViSP ($\sim10$~s). Under this $\sim10$~s cadence, DKIST ViSP exhibits a sensitivity better than $\sim10^{-3}I_c$, thereby meets the requirement discussed in the previous paragraph for active regions. Besides, for the lifetime of reconnection events $\sim200$~s, the scanning range could be $\sim1$~Mm, which is much larger than the typical spatial scale of Stokes features ($\sim0.5$~Mm). Thus, the result of this study should be appropriate to be compared with observations and similar features could be captured in observations.

\subsection{Comparison to Previous Studies on Spectroscopic Signatures of Chromospheric Reconnection} \label{sec:dis:prestudy}

Numerous studies have observed candidates of reconnection events on \CaIR\ profile. Many of them \citep[e.g.,][]{Shetye18,Gosic2018,Vissers19} were focusing on Ellerman bombs with wing enhancement on H$\alpha$ and \CaIR. \citet{Shetye18} also reported some \CaIR\ profiles with weaker absorption on the core, suggesting reconnections in the photosphere with modest heating on the lower chromosphere, which occurs lower than the reconnection events discussed in this work. 

\citet{Gosic2018} and \citet{Vissers19} show \CaIR\ profiles with double peaks around the line core. Instead of interpreted as the high temperature bidirectional outflow, these features can arise when reconnection heats layers below the line core formation height (see Sec.~\ref{sec:res:rxe1:wingsig}). In this scenario, a hot, dense lower layer produces broad emissions, while the overlying cool mid-chromosphere imposes a narrow absorption core, yielding two emission peaks around the absorption line core. 

From inversion, \citet{Vissers19} provides that the heating occurs around $\hbox{log $\tau_{500}$}\sim-$3, beneath the typical formation layer of \CaIR\ line core, which is consistent with the explanation in this work. The temperature increases from $\sim5$~kK to $\sim7$~kK also fits well with our result. Their inferred LOS velocities, however, reach $\sim20$~km~s$^{-1}$ (versus $\sim5$~km~s$^{-1}$ in this work), and their line core widths ($\sim0.8$~\AA\ vs. $\sim0.4$~\AA) are broader. These differences might reflect microturbulence or unresolved small-scale motions in the observed events.

This work focuses on features around the \CaIR\ line core, such as core emissions, which also appears in previous observational studies of reconnections. \citet{Reid17} observed core emission profiles with wing enhancement on its footpoint. Their spectral inversions revealed temperature increases of $1\text{--}2$~kK at $\hbox{log $\tau_{500}$}\sim-$3.5, and bidirectional LOS velocities of $\sim10$~km~s$^{-1}$. Compared to previous studies of Ellerman bombs, this event occurs at a higher height, which is similar to our result. Still, the reconnection height is lower than the typical mid-chromosphere ($\hbox{log $\tau_{500}$}\sim-$5), which likely accounts for the lower outflow velocities compared to our results. In addition, footpoints with wing enhancement also appear in our result, supporting a common underlying mechanism. 

\citet{Kuckein18} also observed \CaIR\ line core emission with temperature increase of $\sim600$~K and LOS speed of $<1$~km~s$^{-1}$ around the mid-chromosphere. The accompanied coronal brightenings indicate reconnections above the chromosphere as microflares, preventing the direct comparison to our result in the mid-chromosphere. Nevertheless, their \CaIR\ emissions last at least $1.5$~min, slightly longer than that in our result ($\sim50$~s). The microflare lifetime ($3$~min) is comparable to the $\sim200$~s duration of our Doppler shift features, which we interpret as the reconnection timescale.

\citet{NS2017} performed a two-dimensional three-component simulation to reproduce surges from chromospheric magnetic reconnection. Their results show the upward outflow to be both faster and more spatially extended than the downward outflow, which agrees well with our result in Sec.~\ref{sec:res:rxe1}. In observation,\citet{Matsumoto08} and \citet{Vissers19} report similar asymmetries of bipolar outflow, which is a consequence of density stratification.

\citet{DB2021} discussed a reconnection event in observation with \CaIR\ polarization signals well above the noise. One example profile shows core emission as in our result. The asymmetry of bipolar reconnection outflow seems to appear around X-point on inverted LOS velocity at $\log_{10}\tau=-4.0$ in both spatial distribution and maximum speed. Polarity of longitudinal magnetic field is also reversed across the X-point at $\log_{10}\tau=-4.0$ as our result. As for the linear polarization, due to the differences on the focused physics, the large range of the colorbar prevented us from checking the LP reduction. Still, we found the linear polarization signal around X-point is lower than $0.1\%I_c$, which might show the potential of LP reduction. The temperature increase of over $2000$~K and the reconnection field strength of $\sim200$~G indicate a much stronger reconnection event than our results.

\citet{daSS22} identified a current sheet from Stokes inversions and extrapolations. The enhanced temperature and the \CaIR\ core emissions point to heating in the mid-chromosphere. However, strong total linear polarization signals appear between the bipolar field patches, which is contrast to the LP reduction from our result, therefore we could infer that no reconnection occurs in the \CaIR\ sensitive layer. This conclusion agrees with the compared simulation in their study, in which the current sheet is above the chromosphere. 

\citet{DB2021} reports mean radiative losses from \Ca3 atoms of $78$~kW~m$^{-2}$ in their most extreme reconnection region, whereas \Citet{daSS22} finds only $6$~kW~m$^{-2}$. 
Applying the method from \citet{DB2021} to our result yields mean losses of $20\text{--}30$~kW~m$^{-2}$ over a $1$~Mm scale and $50$~s interval. The peak radiative loss reaches $187$~kW~m$^{-2}$ at the pixel scale ($\sim20$~km), but falls to $\sim50$~kW~m$^{-2}$ if averaged over $200$~km. Our result is lower than the intense event in \citet{DB2021}, but higher than \citet{daSS22}, where heating is confined above the chromosphere.

In this work, magnetic islands are reproduced as low-speed huge ones and high-speed tiny ones. Synthetic Stokes~$I$ profiles from both types show emissions and broad profile as previous studies \citep{Innes15,Rouppe17} probably due to the penetration of outflow into these magnetic islands (Sec.~\ref{sec:res:rxe2:mgilad}). \citet{Innes15} discussed several blobs with relative low velocities and $>1$\arcsec\ scale, which may corresponds to the low-speed huge magnetic islands in our work. However, their much higher Alfv\'en speed imply reconnection at higher atmospheric heights, preventing further comparision. \citet{Rouppe17} and \citet{DB2021} reported numerous 100--200~km blobs moving at Alfv\'enic speed with 10--20~s lifetime, which is consistent with the high-speed tiny ones in our result. Over the final 100~s of Sec.~\ref{sec:res:rxe2:mgilad}, we identify 5 tiny islands, which slightly exceeds the rate of one blob per 30~s found in \citet{DB2021}.

\subsection{Comparison of the Stokes Features between Magnetic Reconnections and Shock Waves} \label{sec:dis:shk}
Shocks ubiquitously appear in the solar atmosphere. The acoustic waves generated by the convection motion could steepen into shocks after leaked into the photosphere. Regarding the density, temperature, and velocity jump between each side of shock fronts, shocks could lead to observable Stokes~$I$ features such as chromospheric grains \citep{Carlsson92} and umbral flashes \citep{Kneer81}. These Stokes~$I$ features show similarities with those of reconnections, thus additional non-negligible differences are required to discriminate reconnections from shocks in observation. Following the shock identification method from \citet{Wang2021} which mainly detect the velocity jump, we analyze the Stokes signals of the shocks identified from the simulation result. Here we simply compare the main characteristics which also appear in previous studies \cite[e.g.,][]{Felipe14,dela15,Mathur2022,JMG23}. The detail will be discussed in our following papers as this work focuses on magnetic reconnections in this paper.

On the map for the Stokes~$I$, magnetic reconnection current sheets could show two kinds of emission components. One consists of a blue-shifted emission component due to the upward reconnection outflow and an absorption component around the line core attributed to the background atmosphere (e.g., $x=1.3$~Mm on Fig.~\ref{fig:mrx1:viewI}e). The other one contains an emission component around the line core, which often shows slightly red shift due to the suppressed downward outflow (e.g., $x=1.6$~Mm on Fig.~\ref{fig:mrx1:viewI}j). Similar to the upward reconnection flow, the Stokes~$I$ profile of shocks mostly shows emission on the blue side \citep{JMG23}, but the Doppler shift \cite[e.g., $\sim0.25$~\AA\ in][]{Mathur2022} is much lower than that of reconnections ($\sim1$~\AA). Also, the blue-shifted component is isolated from the line core component in the reconnection case. However, if the reconnection current sheet is more inclined, the Doppler blue shift could be much lower. Moreover, outflows are accelerated inside the reconnection current sheet (Fig.~\ref{fig:mrx1:viewI}d\&i), if observed around the center of the reconnection current sheet, the upward outflow component could be close to the line core component, thus becomes similar to the Stokes~$I$ profile of shocks.

More differences appear on the polarization signal. In magnetic reconnection, the Stokes~$V$ profiles show immediate sign reversal around the current sheet, while Stokes~$Q$ reduction appears around the reconnection current sheet. As for the shocks, \citet{Felipe14} and \citet{JMG23} discussed the Stokes~$V$ signal of umbral flashes and chromospheric grains in which shocks propagate upward along the field line. Significant Stokes~$V$ signals appear, usually without reversal of field polarity, which could serve as a reliable difference. On the other hand, \citet{Felipe14} suggest the three-lobe features of the Stokes~$V$ profile as the key feature of shocks. However, three-lobe Stokes~$V$ profiles may also appear in reconnections when the upward outflow component is close to the line core component (e.g., $x=1.0$~Mm on Fig.~\ref{fig:mrx1:iqv}). 

\citet{dela15} presented another model of shocks on emergent loops, showing that linear polarization is highly enhanced in the center of the loops. Thus, considerable LP signals could be regarded as a reliable difference compared to reconnections, in which LP reduction appears around the reconnection current sheet. Additionally, as the fields are nearly horizontal around the small-scale loop apex, Stokes~$V$ signals become almost negligible. Meanwhile, vertical field remain strong with different polarities on either side of the loops, leading to strong Stokes~$V$ signals. Therefore, instead of immediate signal jump in the case of reconnections, the Stokes~$V$ signals appear as gradual transitions from one identifiable state to the other state (e.g., sign reversal) with neutral signal regions at the center, representing another noticeable difference.

In addition, some differences appear on timescale. The shock features on \CaIR\ show timescale $\sim50\text{--}100$~s, which indicates the passing time of shock fronts through the sensitive layer of spectrum. As for reconnections, Doppler shift related to reconnection outflow could exist about $\sim200$~s. On the other hand, emission related to reconnections could last for $\sim50$~s, which is much shorter than the outflow lifetime. Regarding the discussion in Section~\ref{sec:dis:prestudy}, the outflow lifetime is more reliable. Thus, the lifetime of shocks could be much shorter than that of the reconnection outflow in our work.

\section{Conclusion} \label{sec:ccls}
We investigated the Stokes features of magnetic reconnection events in the synthesized full Stokes profile of the simulated solar chromosphere. We performed \CaIR\ Stokes profile synthesis with grid spacing $\sim40$~km and wavelength interval $50$~m\AA\ on realistic radiative MHD simulations with grid spacing $\sim20$~km. Several reconnection events appear within $4000$~s after a magnetic flux is imposed into a well-relaxed unipolar atmosphere, with temperature up to $\sim7$~kK and outflow velocity up to $\sim35$~km~s$^{-1}$. 

Several kinds of Stokes features regarding the temperature enhancement, reconnection outflow, magnetic islands, and magnetic topology are identified. Blue Doppler shift features, up to $\sim1$~\AA\, appear in every reconnection event with acceleration trace connected to the current sheet center, as the Stokes features of the upper outflows. On the other hand, downward outflows are usually suppressed or even blocked due to the dense plasma below the current sheets. Thus, the reconnection outflows lead to asymmetric Doppler shift on Stokes profile, in which blue Doppler shift could reach $\sim1$~\AA\ while red Doppler shift might only reach $\sim0.5$~\AA\ in the vertical case. These Doppler shift features usually last for $\sim200$~s in our result. Emission features appear during these reconnection events when temperature rises to $>7$~kK. These emission features usually last for $\sim50$~s, which is much shorter than the whole reconnection process.


In the polarization profile, many features appear as the manifestation of magnetic field topology. Sign reversal features on Stokes~$V$ profile could be located around the reconnection current sheets, reflecting bipolar vertical magnetic field topology. Stokes~$Q$ features around the reconnection sheet are weaker than surroundings, named as Stokes~$Q$ reduction/LP reduction. This signal reduction co-exists with the intensity features of magnetic reconnection sheets. We suggest that this reduction on the linear polarization signal indicates that plasmas are concentrated around the center of the reconnection current sheet where the magnetic field are relatively weak.

Magnetic islands may appear during the reconnection process with both intensity and polarization signal enhancement. We reproduced several tiny high-speed magnetic islands with Alfv\'enic speed and some huge slow-moving magnetic islands. Tiny magnetic islands lead to signal enhancement mainly on Stokes~$I$ and Stokes~$V$ profile, while in the case of huge magnetic islands, all the components of Stokes profile show considerable enhancement, which may overlap with signal from magnetic reconnection current sheets. 

In addition, regarding the rapid upward motion and shock heating, shock waves can also lead to emission and high blue Doppler shift, which may be similar to the Stokes~$I$ feature of magnetic reconnection. However, features in both linear and circular polarization show noticeable differences between shocks and reconnections. Therefore, polarization signals may serve as a reference to discriminate between reconnections and shocks. We conclude that both linear and circular polarization signals may reveal the distinctive physical mechanisms in reconnections, and enhance the understanding of magnetic reconnection in observations.

\begin{acknowledgments}
This work was supported by FoPM, WINGS Program, the University of Tokyo. T. Yokoyama, S. Toriumi, and M. Kubo are supported by JSPS KAKENHI Grant No. JP20KK0072. Numerical computations and analyses were carried out on Cray XC50 and analysis servers at the Center for Computational Astrophysics, National Astronomical Observatory of Japan.
\end{acknowledgments}

\bibliographystyle{aasjournal}
\bibliography{draft}

\end{document}